\documentclass[showpacs,aps,prd,nofootinbib,amsmath,amssymb,superscriptaddress,twocolumn,10pt]{revtex4}

\usepackage{txfonts}
\usepackage{graphicx}
\usepackage{dcolumn}
\usepackage{bm}
\usepackage{amssymb}
\usepackage{latexsym}
\usepackage{booktabs}
\usepackage[colorlinks=true, linkcolor=red, citecolor=blue]{hyperref}

\newcommand{\n}{\nonumber}

\begin{document}

\title{Reexamination of inflation in noncommutative space-time after Planck results}

\author{Nan Li}
\email{linan@mail.neu.edu.cn}
\affiliation{College of Sciences, Northeastern University, Shenyang 110004, China}
\author{Xin Zhang~\footnote{Corresponding author}}
\email{zhangxin@mail.neu.edu.cn}
\affiliation{College of Sciences, Northeastern University, Shenyang 110004, China}
\affiliation{Center for High Energy Physics, Peking University, Beijing 100080, China}

\begin{abstract}
An inflationary model in the framework of noncommutative space-time may
generate a nontrivial running of the scalar spectral index, but
usually induces a large tensor-to-scalar ratio simultaneously. With
the latest observational data from the Planck mission, we reexamine
the inflationary scenarios in a noncommutative space-time. We find that
either the running of the spectral index is tiny compared with the recent
observational result, or the tensor-to-scalar ratio is
too large to allow a sufficient number of $e$-folds. As
examples, we show that the chaotic and power-law inflation models
with the noncommutative effects are not favored by the current
Planck data.
\end{abstract}


\maketitle

\section{Introduction} \label{sec:intro}

Inflationary paradigm~\cite{Guth,Linde,Steinhardt}, which
postulated an epoch of accelerated expansion in
the very early Universe, has become one the most successful
branches in modern cosmology. It not only eliminated a number of
long-standing cosmological puzzles, such as the horizon, flatness,
entropy, and monopole problems, but also provided a causal
explanation of the primordial fluctuations. When the scales of the
initial perturbations exceeded the Hubble radius, they lost physical
correlations and got frozen. Later on, in the radiation- and
matter-dominated eras, these inhomogeneities reentered the Hubble
radius, seeding the large-scale structures in the Universe. The most important
example is the anisotropies of the cosmic microwave background
(CMB), which encoded the earliest and cleanest information from the
cosmological inflation.

In the last decade, the observations of the CMB, especially from the
Wilkinson Microwave Anisotropy Probe (WMAP)~\cite{WMAP}, have found
enough evidence to support the predictions of the inflationary
cosmology, which asserted that our Universe is spatially flat, with
adiabatic and almost Gaussian primordial fluctuations, described by
a nearly scale-invariant power spectrum. Very recently, the Planck
mission released data, showing that the exact scale-invariant Harrison--Zel'dovich
spectrum~\cite{Harrison,Zel'dovich} is ruled out at over $5\sigma$
confidence level (CL) by the Planck temperature data combined with
the WMAP large-angle polarization (WP)
data~\cite{Planck16,Planck22}. The scalar spectral index measured by the Planck+WP
data is $n_{\rm s}=0.9603\pm 0.0073$, which is very useful
for constraining inflationary models;  when the tensor
component is included, the spectral index is not significantly
changed, $n_{\rm s}=0.9624\pm 0.0075$. Besides, the Planck+WP data also set a
tighter upper bound on the tensor-to-scalar ratio, $r<0.12$ (at
the pivot scale $k_\ast=0.002$~Mpc$^{-1}$). It is shown in
Ref.~\cite{Planck22} that the space for the allowed standard inflationary
models is shrunk (potentials with $V_{\phi\phi}<0$ are preferred), and some
models cannot provide a good fit to the Planck data.

The running of the scalar spectral index, ${\rm d}n_{\rm s}/{\rm
d}\ln k$, is also measured by the Planck mission. It is found by the
Planck+WP data that ${\rm d}n_{\rm s}/{\rm d}\ln k=-0.0134\pm0.0090$ (at
the pivot scale $k_\ast=0.05$~Mpc$^{-1}$), which is negative at
1.5$\sigma$ level; at the same time, we have $n_{\rm s}=0.9561\pm
0.0080$~\cite{Planck22}. Though the evidence of a negative value for
${\rm d}n_{\rm s}/{\rm d}\ln k$ of ${\cal O}(10^{-2})$ is not
statistically significant enough, such a result is still interesting
for the physics of inflation, since the typical slow-roll
inflationary models can only generate the running of
${\cal O}(10^{-3})$. Obviously, if the result of a large negative
running of the spectral index is confirmed, a new window into
physics of inflation will be opened. So it is of interest to design
inflationary models that predict a negative running of ${\cal
O}(10^{-2})$ with an acceptable $n_{\rm s}$ and number of $e$-folds.

A way of realizing a large negative running is to consider a
slow-roll inflation in a noncommutative space-time. Though it is
still hard to construct a realistic noncommutative inflationary
model, a toy model~\cite{Brandenberger} has been extensively
studied~\cite{HuangLi1,HuangLi2,HuangLi,Huang,Zhang:2006ta,Huang:2006zu,Zhang:2006cc,Cai:2007xr,Cai:2007et,Xue:2007bb},
and was especially examined with the WMAP 3-yr
results~\cite{Huang,Zhang:2006ta,Huang:2006zu,Zhang:2006cc}. Since
there was a strong degeneracy between $n_{\rm s}$ and ${\rm d}n_{\rm
s}/{\rm d}\ln k$~\cite{wmap3}, the WMAP data favored a blue spectrum
and a large negative running, with a large tensor-to-scalar ratio.
It was shown that the WMAP results could be nicely explained by the
noncommutative inflationary
models~\cite{Huang,Zhang:2006ta,Huang:2006zu,Zhang:2006cc}. However,
the Planck mission now places much tighter constraints on the
primordial power spectrum, preferring a red spectrum, with a
negative running of ${\cal O}(10^{-2})$ and a small tensor-to-scalar
ratio. Therefore, it is worthy of reexamining the status of the
noncommutative inflationary models in light of the latest
observational data from the Planck mission~\cite{Planck16,Planck22}.
This is the main purpose of the present paper.

This paper is organized as follows. First, in Sect.~\ref{sec:non},
we briefly review the inflationary cosmology in noncommutative
space-time, list the predictions for the primordial parameters,
and discuss the possibility to realize a negative
running of the spectral index in this framework. With these general preparations, in
Sect.~\ref{sec:scenarios}, we move on to investigate in detail two
specific inflationary scenarios, the chaotic and the power-law
inflation models. We find that the noncommutative effects may give
rise to a red power spectrum and a negative running in these models,
but always induce a too large tensor-to-scalar ratio at the same time,
so cannot thoroughly match the Planck data. Finally, we conclude in
Sect.~\ref{sec:sum}.

\section{Inflation in noncommutative space-time} \label{sec:non}

A noncommutative space-time emerges as a natural consequence from
string theory~\cite{Yoneya,Li}. For the noncommutative inflationary
models, we refer the reader to Ref.~\cite{HuangLi} and the references
therein. In any physical process, the noncommutation of time and
space coordinate implies a new uncertainty relation,
\begin{eqnarray}
\Delta t\Delta x\geq l_{\rm s}^2=\frac{1}{M_{\rm s}^2}, \n
\end{eqnarray}
where $t$ and $x$ are the physical time and space coordinate,
$l_{\rm s}$ is the string length, which is the fundamental degree of
freedom in string theory, and $M_{\rm s}:=l^{-1}_{\rm s}$ is the
string mass scale. In general, if $M_{\rm s}$ is close to the energy
scale at which inflation took place, the noncommutative effects will
accordingly impact the details of cosmological inflation. Consequently,
observable imprints may be left on the late-time evolution of the
Universe, e.g., the CMB angular power
spectrum~\cite{Brandenberger,HuangLi1,HuangLi2}.

\subsection{Noncommutative effects in inflation} \label{sec:brief}

In this paper, we restrict our discussion in the single-field
slow-roll inflationary models, and explore the influences from the
noncommutative effects on it. In this framework, the evolution of
the homogeneous and isotropic background and the Klein--Gordon
equation for the inflaton field will not be changed in
noncommutative space-time,
\begin{eqnarray}
&&H^2=\left(\frac{\dot{a}}{a}\right)^2=\frac{1}{3m_{\rm P}^2}\left[\frac12\dot{\phi}^2+V(\phi)\right], \n\\
&&\ddot{\phi}+3H\dot{\phi}+V_{\phi}(\phi)=0, \n
\end{eqnarray}
where $a(t)$ is the scale factor, $H:=\dot{a}/a$ is the Hubble
expansion rate, $\phi$ is the inflaton field, $V(\phi)$ is its
potential (with the subscript $_{\phi}$ being the partial derivative
with respect to $\phi$), and $m_{\rm P}:=1/\sqrt{8\pi G}=2.44\times
10^{18}$~GeV is the reduced Planck mass. For simplicity, we assume
that the Universe is spatially flat. In the slow-roll approximation, the inflaton field slowly evolves
down its potential, and three slow-roll parameters may thus be
introduced as
\begin{eqnarray}
\epsilon_V:=\frac{m_{\rm P}^2}{2}\left(\frac{V_\phi}{V}\right)^2,
\quad \eta_V:=m_{\rm P}^2\frac{V_{\phi\phi}}{V}, \quad \xi_V:=m_{\rm
P}^4\frac{V_\phi V_{\phi\phi\phi}}{V^2}. \n
\end{eqnarray}
Hence, the nearly exponential expansion of the Universe is achieved
under the conditions $\epsilon_V\ll 1$ and $|\eta_V|\ll 1$.

Despite the unchanged evolution of the background space-time, the
influences from the noncommutative effects show up in the evolutions
of the cosmological perturbations and their power spectra. To
express these more explicitly, a new time coordinate, namely, the
modified conformal time $\tilde{\eta}$, is introduced. In this way,
the equation of motion for the Fourier mode of the comoving
curvature perturbation $\mathcal{R}_k$ reads
\begin{eqnarray}
\mathcal{R}_k''+2\frac{z_k'}{z_k}\mathcal{R}_k'+k^2\mathcal{R}_k=0, \label{R}
\end{eqnarray}
where we denote the derivative with respect to $\tilde{\eta}$ by $'$, and
\begin{eqnarray}
\frac{{\rm d}\tilde{\eta}}{{\rm
d}\eta}&=&\left(\frac{\beta_k^-}{\beta_k^+}\right)^{1/2}, \quad
z_k(\tilde{\eta})=\frac{a\dot{\phi}}{H}\left(\beta_k^+\beta_k^-\right)^{1/4}, \n\\
\beta^+_k&=&\frac12\left[a^2(\eta+l_{\rm s}^2k)+a^2(\eta-l_{\rm s}^2k)\right], \n\\
\beta^-_k&=&\frac12\left[\frac{1}{a^2(\eta+l_{\rm s}^2k)}+\frac{1}{a^2(\eta-l_{\rm s}^2k)}\right], \n
\end{eqnarray}
where $\eta$ is the conformal time. The deviation of the equation of
motion for $\mathcal{R}_k$ in noncommutative space-time from that in
the ordinarily commutative space-time is encoded in the terms
$\beta_k^+$ and $\beta_k^-$, and furthermore can be characterized by
a noncommutative parameter $\mu$~\cite{HuangLi},
\begin{eqnarray}
\mu:=\left(\frac{Hkl_{\rm s}}{a}\right)^2=\left(\frac{Hk}{aM_{\rm
s}}\right)^2. \n
\end{eqnarray}

Solving Eq.~(\ref{R}), we arrive at the following basic parameters for the
power spectrum of the comoving curvature perturbation (for detailed
derivations of the following results, we refer the reader to
Ref.~\cite{HuangLi}):
\begin{enumerate}
\item the power spectrum,
\begin{eqnarray}
\mathcal{P}_\mathcal{R}=\frac{k^3}{2\pi^2}|\mathcal{R}_k|^2=\frac{V}{24\pi^2m_{\rm
P}^4\epsilon_V}(1+\mu)^{2\eta_V-6\epsilon_V-4}, \label{power}
\end{eqnarray}
\item the scalar power spectral index,
\begin{eqnarray}
s=n_{\rm s}-1:=\frac{{\rm d}\ln \mathcal{P}_\mathcal{R}}{{\rm d}\ln
k}=2\eta_V-6\epsilon_V+16\epsilon_V\mu, \label{index}
\end{eqnarray}
\item the running of the scalar spectral index,
\begin{eqnarray}
\alpha_{\rm s}:=\frac{{\rm d}n_{\rm s}}{{\rm d}\ln
k}=-24\epsilon_V^2+16\epsilon_V\eta_V-2\xi_V-32\epsilon_V\eta_V\mu,
\label{running}
\end{eqnarray}
\item the tensor-to-scalar ratio,
\begin{eqnarray}
r=16\epsilon_V. \label{ratio}
\end{eqnarray}
\end{enumerate}
From Eqs.~(\ref{power})--(\ref{ratio}), we may clearly observe the
noncommutative effects, parameterized by $\mu$, on the usual results
of the inflationary cosmology. In the limit of vanishing
noncommutative effects, $\mu\rightarrow 0$, all these results
automatically reduce to the usual expressions in the standard slow-roll
inflationary paradigm.

\subsection{Running of the spectral index in noncommutative inflation}
\label{sec:running}

Now we take the running of the scalar spectral index $\alpha_{\rm s}$
as an example to give a very general discussion of the noncommutative effects
on the primordial parameters. From Eqs.~(\ref{index}) and (\ref{ratio}),
we have $\eta_V=\frac12(\frac38r+s-r\mu)$. Therefore, $\alpha_{\rm
s}$ can be reexpressed as
\begin{eqnarray}
\alpha_{\rm
s}=r^2\mu^2-\left(\frac78r+s\right)r\mu+\frac{3}{32}\left(r+\frac{16}{3}s\right)r-2\xi_V. \n
\end{eqnarray}
Usually, it is not easy to construct a large $\xi_V$ in the known
inflationary models, so in the following we neglect $\xi_V$ for
simplicity. Thus,
\begin{eqnarray}
\alpha_{\rm s}=r^2\mu^2-\left(\frac78r+s\right)r\mu+\frac{3}{32}\left(r+\frac{16}{3}s\right)r. \label{alpha}
\end{eqnarray}
Equation~(\ref{alpha}) indicates that $\alpha_{\rm s}$ is a
quadratic function of the noncommutative parameter $\mu$ and
is thus possible to be negative, as it is
straightforward to see that the discriminant for Eq.~(\ref{alpha}) is positive definite,
\begin{eqnarray}
\Delta=r^2\left[\frac38r^2+\left(\frac18r-s\right)^2\right]>0. \n
\end{eqnarray}
We classify $\alpha_{\rm s}$ into the following three cases:
\begin{enumerate}
\item For $s>0$, $\alpha_{\rm s}$ is negative, if $\mu_-<\mu<\mu_+$,
with $\mu_{\pm}=\frac{1}{2r^2}\left[r\left(\frac78r+s\right)\pm\sqrt{\Delta}\right]$.
\item For $s<0$ and $r<-\frac{16}{3}s$, $\alpha_{\rm s}$ is negative,
if $0<\mu<\mu_+$.
\item For $s<0$ and $r>-\frac{16}{3}s$, $\alpha_{\rm s}$ is negative, if $\mu_-<\mu<\mu_+$.
\end{enumerate}
From these analyses, we can conclude that there exists the
possibility of a negative running of the spectral index in the
noncommutative inflationary models. This is consistent with
the discussions in Ref.~\cite{Huang}.
We should admit that $\alpha_{\rm s}$ may also be negative in the
inflationary models in commutative space-time. As can be seen from Eq.~(\ref{alpha}),
$\alpha_{\rm s}=\frac{3}{32}(r+\frac{16}{3}s)r$, when $\mu=0$. Therefore, $\alpha_{\rm s}$
is negative, if $s<0$ and $r<-\frac{16}{3}s$. But with a new parameter $\mu$
from the noncommutative effects, $\alpha_{\rm s}$ may be negative
for much larger ranges of $r$ and $s$.

In Refs.~\cite{Huang,Zhang:2006ta,Huang:2006zu,Zhang:2006cc}, the
noncommutative inflationary models were compared to the WMAP 3-yr
results. However, at that time, the measurements of the spectral
index $n_{\rm s}$ and its running $\alpha_{\rm s}$ were still
imprecise, $n_{\rm s}=1.21^{+0.13}_{-0.16}$ and $\alpha_{\rm
s}=-0.102^{+0.050}_{-0.043}$~\cite{wmap3}. The upper bound for the
tensor-to-scalar ratio $r$ was even looser, $r<1.5$ at the $95\%$ CL
(WMAP data only)~\cite{wmap3}. We see that a blue spectrum with a
large negative running was favored, and large tensor
components were consistent with these data. In Ref.~\cite{Huang}, it
was shown that the noncommutative chaotic inflation models could
realize a considerable running index for the $\phi^2$ and
$\phi^4$ cases, but a low number of $e$-folds, say, $N\sim 14$,
is required. Subsequently, it was pointed out in
Ref.~\cite{Zhang:2006ta} that the noncommutative chaotic inflation
models could provide a large negative running of the spectral index
within a reasonable range of the number of $e$-folds, provided that the power $n$ in
the potential $V(\phi)\sim\phi^n$ was enhanced roughly to $n\sim 12$--$18$.
In addition, for the Kachru--Kallosh--Linde--Maldacena--McAllister--Trivedi brane inflation
model~\cite{KKLMMT}, the noncommutative effects not only could nicely explain the large
negative running of the spectral index, but also significantly
relaxed the fine-tuning for the parameter $\beta$~\cite{Huang:2006zu,Zhang:2006cc} (see also
Refs.~\cite{Huang:2006ra,Ma:2008rf,Ma:2013xma} for the fine-tuning
problem).

It has been shown by the Planck mission that compared to previous experiments,
the Planck data make the space of possible inflationary models
narrower. In particular, a preference for a negative running of $n_{\rm
s}$ at modest statistical significance is still held, 
which has been challenging the design of slow-roll
inflationary models. Thus, it is of interest to rescrutinize
the noncommutative inflationary models with the current Planck data. We wish to
know if the noncommutative mechanism still works when
confronting the Planck data. At least, we should make certain
whether the noncommutative inflationary models help to provide a better fit to the data
or make the situation worse.

In what follows, we quote the fit results for the $\Lambda$ cold dark matter
($\Lambda$CDM) model, with a two-parameter extension, i.e., the
$\Lambda$CDM+$r$+$\alpha_{\rm s}$ model, from the Planck data combined with the
WMAP large-scale polarization data and the baryon acoustic oscillation
(BAO) data (henceforth, Planck+WP+BAO)~\cite{Planck22},
\begin{eqnarray}
n_{\rm s}&=&0.9607\pm0.0063 \quad (68\%~{\rm CL}), \n\\
\alpha_{\rm s}&=&-0.021^{+0.012}_{-0.010} \quad (68\%~{\rm CL}), \n\\
r&<&0.25 \quad (95\%~{\rm CL}), \label{Planck}
\end{eqnarray}
which are given at the pivot scale $k_\ast=0.05~{\rm Mpc}^{-1}$.
Comparing to the WMAP results~\cite{wmap3}, we find that the spectral index
$n_{\rm s}$ in this case is less than 1 rather than greater than 1
(red, not blue spectrum). This is not a good signal for the
noncommutative effects, as from Eq.~(\ref{index}) we see that
the noncommutative effects always make the power spectrum more blue,
since $\epsilon_V$ must be positive. Also, we find that in this case a
negative running of the spectral index is favored at about 2$\sigma$
level. From Eq.~(\ref{running}), we observe that, if $\eta_V$ is
positive, the noncommutative effects will help to realize a
negative running; otherwise, the role of the noncommutative effects is to prevent a negative running.
This is a bad hint for the noncommutative inflation,
because the Planck data prefer potentials
with $V_{\phi\phi}<0$, i.e., $\eta_V<0$. The constraint on the tensor-to-scalar ratio is
relaxed compared to the case with no running, due to an
anti-correlation between $r$ and $\alpha_{\rm s}$ in the Planck+WP+BAO
data. But such an upper bound for $r$ is still fairly strict for
constraining noncommutative inflation.

We will show in the following that the noncommutative effects may
induce a negative running of the spectral index $\alpha_{\rm s}$
but will simultaneously generate a too-large
tensor-to-scalar ratio $r$. In fact, as shown immediately in
Sect.~\ref{sec:scenarios}, the balance between $\alpha_{\rm s}$ and $r$ can never be
achieved in the usual inflationary models in noncommutative space-time.
As a result, the noncommutative
inflationary models cannot survive under the strict constraints in
Eq.~(\ref{Planck}) and are thus excluded by the current data.

Actually, several months before the Planck data were released, the South Pole Telescope (SPT)~\cite{SPT}
and the Atacama Cosmology Telescope (ACT)~\cite{ACT} Collaborations released their CMB data,
presenting accurate CMB angular power spectra which in fact extend to smaller angular scales.
The SPT and ACT results are consistent with the Planck results.
They both favor a red tilt of the primordial power spectrum.
When considering the running of the spectral index, the SPT data combined with
the WMAP 7-yr data give $\alpha_{\rm s}=-0.024\pm 0.011$ (at the pivot scale
$k_\ast=0.025$~Mpc$^{-1}$), favoring a negative running at the 2.2$\sigma$ level.
The preference for $\alpha_{\rm s}<0$ strengthens to 2.7$\sigma$ for the combination
of SPT+WMAP-7+BAO+$H_0$.
The ACT data are, however, consistent with no running of the spectral index,
$\alpha_{\rm s}=-0.004\pm0.012$ (ACT+WMAP-7), at the pivot scale $k_\ast=0.015$~Mpc$^{-1}$.
For the $\Lambda$CDM+$r$ cosmology, the ACT Collaboration obtains
$r<0.16$ (95\% CL; ACT+WMAP-7+BAO+$H_0$).
These observational results are, in fact, also sufficient to exclude the noncommutative
inflation model. Since the Planck data are the latest observational data, in this paper
we only use the Planck results to make a detailed analysis but do not duplicate the similar
discussion using the SPT and ACT results. However, we should keep in mind that the
SPT and ACT results would give the same conclusion even if we did not know the
Planck results.

\section{Noncommutative effects in specific inflationary scenarios} \label{sec:scenarios}

In this section, some most frequently discussed inflationary scenarios
are inspected and reexamined in detail in noncommutative space-time.
We find that the noncommutative inflationary models are in serious conflict
with the current Planck+WP+BAO data.

\subsection{Chaotic inflation scenario} \label{sec:chaotic}

The chaotic inflation scenario~\cite{Lindechaotic} is a typical
large-field model, in which the potential takes the form
$V(\phi)\sim\phi^n$, and the inflaton field $\phi$ is usually
displaced from the minimum of $V(\phi)$ by an amount of the order of
$m_{\rm P}$. The calculations of the slow-roll parameters in this
model are a standard procedure in inflationary cosmology,
\begin{eqnarray}
\epsilon_V=\frac{n}{4N}, \quad \eta_V=\frac{n-1}{2N}, \quad \xi_V=\frac{(n-1)(n-2)}{4N^2}, \n
\end{eqnarray}
where $N$ is the number of $e$-folds of inflation. Furthermore, the
spectral index, the running of the spectral index, and the
tensor-to-scalar ratio in noncommutative space-time are~\cite{HuangLi}
\begin{eqnarray}
s=n_{\rm s}-1&=&\left(\mu-\frac{n+2}{8n}\right)r, \label{schaotic} \\
\alpha_{\rm s}&=&-\frac{n-1}{4n}\left[\mu+\frac{n+2}{8n(n-1)}\right]r^2, \label{alphachaotic}\\
r&=&\frac{4n}{N}. \label{rchaotic}
\end{eqnarray}
From Eqs.~(\ref{schaotic}) and (\ref{alphachaotic}), we get a red
spectral index, if $\mu<\frac{n+2}{8n}$, and a negative definite
running of the spectral index, $\alpha_{\rm
s}<-\frac{n+2}{32n}r^2$. Therefore, in some
sense, the noncommutative effects do help to induce a larger
negative running of the spectral index (compared to that in the
commutative case), but further analyses will reveal that these
noncommutative effects also lead to a too large tensor-to-scalar
ratio unavoidably.

From Eqs.~(\ref{schaotic}) and (\ref{alphachaotic}), we can also
solve $r$ and $\mu$ as the functions of $s$ and $\alpha_{\rm s}$,
\begin{eqnarray}
r&=&\frac{4(n-1)}{n+2}\left[\sqrt{s^2-\frac{2n(n+2)}{(n-1)^2}\alpha_{\rm s}}-s\right], \label{mur}\\
\mu&=&\frac{n+2}{8n}-\frac{(n-1)s}{8n\alpha_{\rm s}}\left[\sqrt{s^2-\frac{2n(n+2)}{(n-1)^2}\alpha_{\rm s}}+s\right]. \label{rmu}
\end{eqnarray}
With the Planck+WP+BAO constraints~\cite{Planck22},
\begin{eqnarray}
s=n_{\rm s}-1=-0.0393\pm0.0063, \quad \alpha_{\rm s}=-0.021^{+0.012}_{-0.010}, \label{salpha}
\end{eqnarray}
it is not difficult to attain the allowed ranges for $r$ and $\mu$.
Moreover, from Eqs.~(\ref{schaotic})--(\ref{rchaotic}), it is
straightforward to figure out the $r$--$n_{\rm s}$ and $\alpha_{\rm
s}$--$n_{\rm s}$ relationships,
\begin{eqnarray}
r&=&\frac{n_{\rm s}-1}{\mu-\frac{n+2}{8n}}, \n\\
\alpha_{\rm s}&=&-\frac{n+2}{32n}r^2-\frac{n-1}{4n}(n_{\rm s}-1)r \n\\
&=&-\frac{n(n+2)}{2N^2}-\frac{n-1}{N}(n_{\rm s}-1). \n 
\end{eqnarray}

Now we pick two most often used cases, the $\phi^2$ and
$\phi^4$ inflation models, as examples.

\subsubsection{$\phi^2$ inflation model}

In this case, $n=2$. From Eqs.~(\ref{schaotic})--(\ref{rmu}), we
have
\begin{eqnarray}
s=\left(\mu-\frac14\right)r, \quad \alpha_{\rm s}=-\frac18\left(\mu+\frac14\right)r^2, \quad r=\frac8N, \label{phi2}
\end{eqnarray}
and
\begin{eqnarray}
r&=&\sqrt{s^2-16\alpha_{\rm s}}-s, \n\\
\mu&=&\frac14-\frac{s}{16\alpha_{\rm s}}\left(\sqrt{s^2-16\alpha_{\rm s}}+s\right). \n
\end{eqnarray}

\begin{figure}[h]
\begin{center}
\includegraphics[width=.85\linewidth]{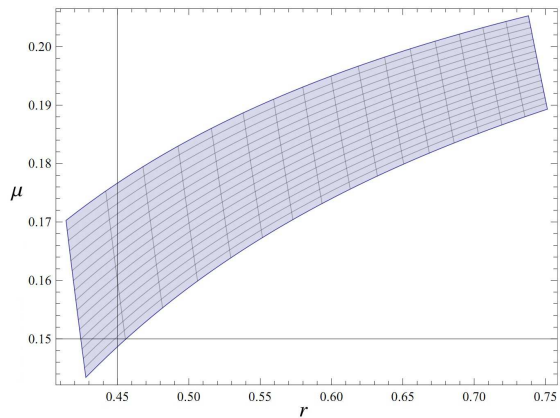}
\caption{The expected region of the tensor-to-scalar ratio $r$ and
the noncommutative parameter $\mu$ in the $\phi^2$ inflation model
under the consideration of the 1$\sigma$ ranges of $n_{\rm s}$ and
$\alpha_{\rm s}$ in Eq.~(\ref{Planck}). We find that the
noncommutative effects lead to a too large tensor-to-scalar ratio,
$0.41<r<0.75$. Here we simply treat $s$ and $\alpha_{\rm s}$ as free
parameters, and the lower bound of $r$ will be even larger, if the
correlation of $s$ and $\alpha_{\rm s}$ is considered.}\label{fig:2}
\end{center}
\end{figure}

The expected region of the tensor-to-scalar ratio $r$ and the
noncommutative parameter $\mu$ in the $\phi^2$ inflation model is
shown in Fig.~\ref{fig:2}. It is clear to see that the noncommutative $\phi^2$
inflation model gives rise to a too large tensor-to-scalar ratio, $r>0.41$,
seriously inconsistent with the Planck+WP+BAO result,
$r<0.25$~\cite{Planck22}. In fact, in Fig.~\ref{fig:2}, we
simply treat $s$ and $\alpha_{\rm s}$ as free parameters, with the
constraints in Eq.~(\ref{salpha}), and neglect their correlation,
while this is already enough for us to conclude that the
noncommutative effects are not favored by the current observations,
because the allowed ranges of $r$ and $\mu$ will be even narrower,
with the correlation of $s$ and $\alpha_{\rm s}$ taken into account.

\begin{figure}[h]
\begin{center}
\includegraphics[width=0.9\linewidth]{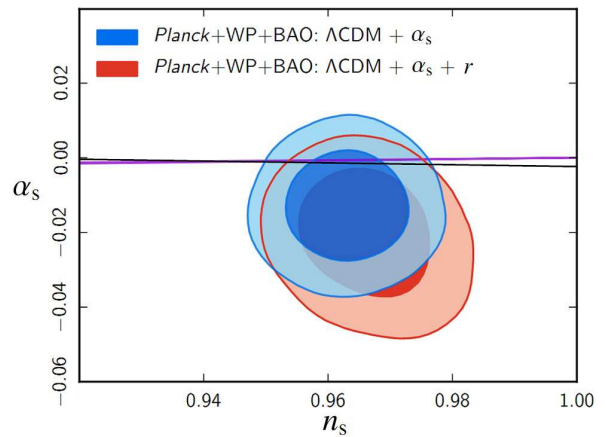}
\caption{The predicted relationship of the running of the spectral
index $\alpha_{\rm s}$ and the spectral index $n_{\rm s}$ (black
line) in the $\phi^2$ inflation model in noncommutative space-time,
with the number of $e$-folds chosen to be $N=40$. Contours ($68\%$
and $95\%$ CLs) in the $\alpha_{\rm s}$--$n_{\rm s}$ plane are from
the joint Planck+WP+BAO constraints; here we use a copy of Fig.~2 in
Ref.~\cite{Planck22}. Blue contours (upper) are for $\Lambda$CDM+$\alpha_{\rm s}$
cosmology, and red contours (lower) are for $\Lambda$CDM+$\alpha_{\rm s}$+$r$
cosmology. For comparison, we keep the purple strip
that shows the prediction for usual chaotic inflation models in
commutative space-time with $50<N<60$. Note that in the usual slow-roll 
chaotic inflation models $\alpha_{\rm s}$ is of order $(1-n_{\rm s})^2$,
and thus under the constraints of current data this purple strip is 
so narrow that it looks like a purple line.
On the contrary to this strip,
the noncommutative effects lead to a more
negative slope for the $\alpha_{\rm s}$--$n_{\rm s}$ relation. We
find a rather small $\alpha_{\rm s}$ in the noncommutative model,
$\alpha_{\rm s}=-0.0005$ at $n_{\rm s}=0.92$ and $\alpha_{\rm s}=-0.0025$
at $n_{\rm s}=1$. For the central value of the spectral index
$n_{\rm s}=0.9607$~\cite{Planck22}, we get $\alpha_{\rm
s}=-0.001518$. These values are tiny compared with the central value
$\alpha_{\rm s}=-0.021$~\cite{Planck22}, indicating that the
noncommutative $\phi^2$ inflation model is not favored by the
current Planck+WP+BAO observations at the $68\%$ CL (red contours). Larger number of
$e$-folds makes things worse, as the slope of the function in
Eq.~(\ref{phi2alpha}) is even smaller.}\label{fig:alphas-ns}
\end{center}
\end{figure}

In order to avoid the problem of a too large tensor-to-scalar ratio,
we find from Eq.~(\ref{phi2}) that the number of $e$-folds must be
large enough, $N>8/0.25=32$. But this causes another problem. From
Eq.~(\ref{phi2}), we have
\begin{eqnarray}
\alpha_{\rm s}=-\frac{4}{N^2}-\frac{n_{\rm s}-1}{N}. \label{phi2alpha}
\end{eqnarray}
We find that the running of the spectral index $\alpha_{\rm s}$ is a
linear function of the spectral index $n_{\rm s}$. However, for a
large enough $N$, the slope of the function will be too small to
produce a negative enough running. Let us choose $N=40$ ($r=0.2$) as
an example to compare with the Planck+WP+BAO results in the
$\alpha_{\rm s}$--$n_{\rm s}$ plane~\cite{Planck22}. Our result is
shown in Fig.~\ref{fig:alphas-ns}. We observe that $\alpha_{\rm
s}=-0.0005$ at $n_{\rm s}=0.92$ and $\alpha_{\rm s}=-0.0025$ at $n_{\rm
s}=1$. This means that with a suitable number of $e$-folds,
although the noncommutative effects may generate a negative running
of the spectral index $\alpha_{\rm s}$, this running is always too
small. A larger $N$ will make the situation worse, as the slope in
Eq.~(\ref{phi2alpha}) becomes even smaller. To see how serious this
trouble is, we pick the central values of $\alpha_{\rm s}$ and
$n_{\rm s}$ from the Planck+WP+BAO data~\cite{Planck22}, $\alpha_{\rm
s}=-0.021$ and $n_{\rm s}=0.9607$, and by solving
Eq.~(\ref{phi2alpha}), we get $N=17.1$, which is far from enough for
a reasonable inflationary model. The above analysis demonstrates that
the noncommutative effects in the $\phi^2$ inflation model cannot
provide useful help in explaining the large negative running of the
spectral index. From Fig.~\ref{fig:alphas-ns}, we see that the
noncommutative $\phi^2$ inflation model (with $r=0.2$) lies outside of
the $1\sigma$ region of Planck+WP+BAO constraint on the
$\Lambda$CDM+$\alpha_{\rm s}$+$r$ model (red contours).

\begin{figure}[h]
\begin{center}
\includegraphics[width=0.9\linewidth,height=0.75\linewidth]{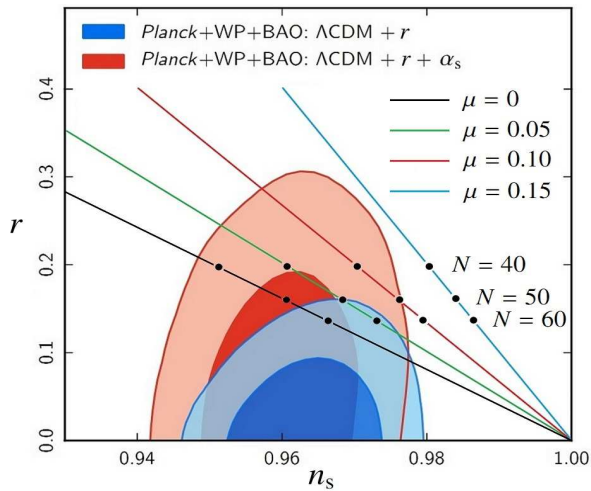}
\caption{The predicted relationship of the tensor-to-scalar ratio
$r$ and the spectral index $n_{\rm s}$ in the $\phi^2$ inflation
model in noncommutative space-time. Contours ($68\%$ and $95\%$ CLs)
in the $r$--$n_{\rm s}$ plane are from the joint Planck+WP+BAO
constraints; here we use a copy of Fig.~4 in Ref.~\cite{Planck22}.
Blue contours (lower) are for $\Lambda$CDM+$r$ cosmology, and 
red contours (upper) are for $\Lambda$CDM+$r$+$\alpha_{\rm s}$ cosmology.
The black line indicates the $\phi^2$ inflation model in commutative
space-time, and the green, red, and blue lines show the $r$--$n_{\rm
s}$ relationships, with the noncommutative parameter $\mu$ taken to
be $0.05$, $0.10$, and $0.15$, respectively. We find that the larger
$\mu$ is, the more inconsistent the noncommutative models are with
the Planck+WP+BAO data. If $r$ is lower than its
upper bound $r<0.25$~\cite{Planck22}, a large enough number of
$e$-folds $N$ is needed. Here we take $N=40$, $50$, and $60$,
respectively. Since the noncommutative $\phi^2$ inflation model predicts a
negligible running, given a reasonable number of $e$-folds,
the theoretical predictions (the green, red, and blue
lines) should be compared to the blue contours in the $r$--$n_{\rm
s}$ plane. We see that the noncommutative $\phi^2$ inflation model
with $\mu\gtrsim 0.05$ is ruled out at the $95\%$ CL. Even if the
comparison is made to the red contours, the model with $\mu\gtrsim
0.05$ is ruled out at the $68\%$ CL, and with $\mu\gtrsim
0.15$ is ruled out at the $95\%$ CL.}\label{fig:r-ns}
\end{center}
\end{figure}

Furthermore, for the $r$--$n_{\rm s}$ relationship, from
Eq.~(\ref{phi2}), we have
\begin{eqnarray}
r=\frac{n_{\rm s}-1}{\mu-\frac14}. \n
\end{eqnarray}
For a positive $r$, $\mu<0.25$. We show the theoretical predictions
of the model in Fig.~\ref{fig:r-ns}, in which we choose $\mu=0.05$,
$0.10$, and $0.15$, respectively, and compare these three cases with
the ordinary $\phi^2$ inflation model in commutative space-time.
Since the noncommutative $\phi^2$ inflation model also predicts a
negligible running, given a reasonable number of $e$-folds, say,
50--60, the theoretical predictions (the green, red, and blue
lines) should be compared to the blue contours in the $r$--$n_{\rm
s}$ plane. We see that the noncommutative $\phi^2$ inflation model
with $\mu\gtrsim 0.05$ is ruled out at the $95\%$ CL. Even if the
comparison is made to the red contours, the model with $\mu\gtrsim
0.05$ is ruled out at the $68\%$ CL, and with $\mu\gtrsim
0.15$ is ruled out at the $95\%$ CL.

In a word, we conclude that the $\phi^2$ chaotic inflation model in
noncommutative space-time is not favored by the current Planck+WP+BAO
results~\cite{Planck22}. If we want to explain the results of
$n_{\rm s}$ and $\alpha_{\rm s}$ in Eq.~(\ref{Planck}), a large
value of $r$ will be inevitable, which severely exceeds the current
observational upper bound. If we abandon explaining the large
running, we will find that the noncommutative effects do not provide
a better fit to the data, but make the situation even worse; in this
case, the noncommutative $\phi^2$ inflation model with $\mu\gtrsim 0.05$
is ruled out by the current data at the $95\%$ CL.

\subsubsection{$\phi^4$ inflation model}

In this case, $n=4$. From Eqs.~(\ref{schaotic})--(\ref{rmu}), we
have
\begin{eqnarray}
s=\left(\mu-\frac{3}{16}\right)r, \quad \alpha_{\rm s}=-\frac{3}{16}\left(\mu+\frac{1}{16}\right)r^2, \quad r=\frac{16}{N}, \n
\end{eqnarray}
and
\begin{eqnarray}
r&=&2\left(\sqrt{s^2-\frac{16\alpha_{\rm s}}{3}}-s\right), \n\\
\mu&=&\frac{3}{16}-\frac{3s}{32\alpha_{\rm s}}\left(\sqrt{s^2-\frac{16\alpha_{\rm s}}{3}}+s\right). \n
\end{eqnarray}

\begin{figure}[h]
\begin{center}
\includegraphics[width=.85\linewidth]{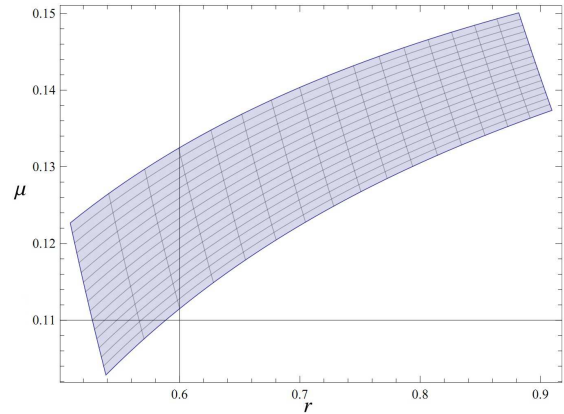}
\caption{The expected region of the tensor-to-scalar ratio $r$ and
the noncommutative parameter $\mu$ in the $\phi^4$ inflation model
under the consideration of the 1$\sigma$ ranges of $n_{\rm s}$ and
$\alpha_{\rm s}$ in Eq.~(\ref{Planck}). The noncommutative effects
lead to a even larger tensor-to-scalar ratio, $0.51<r<0.91$. This
means that the noncommutative $\phi^4$ inflation model is more inconsistent with the
current observational data.}\label{fig:4}
\end{center}
\end{figure}

In fact, the ordinary $\phi^4$ inflation model in commutative space-time has
been ruled out at more than 3$\sigma$ level by the current Planck
data~\cite{Planck22}. Now let us see the status of the
noncommutative case. Similarly as the $\phi^2$ inflation model,
if the values of $n_{\rm s}$ and $\alpha_{\rm s}$ in
Eq.~(\ref{Planck}) are taken, the expected region of the
tensor-to-scalar ratio $r$ and the noncommutative parameter $\mu$ in
the $\phi^4$ inflation model is shown in Fig.~\ref{fig:4}. In this
case, the lower bound of $r$ is even higher, $r>0.51$, in obvious
contradiction with its current observational bound. Thus, for
the $\phi^4$ model, the noncommutative effects give even worse
predictions than those in the $\phi^2$ inflation model.

The considerations on the $\alpha_{\rm s}$--$n_{\rm s}$ and
$r$--$n_{\rm s}$ relationships in the noncommutative $\phi^4$ inflation model
are totally analogous to those in the
$\phi^2$ inflation model, only with more severe discrepancies, so we do not
duplicate our analyses.

\subsection{Power-law inflation scenario} \label{sec:powerlaw}

The power-law inflation scenario~\cite{powerlaw}
is driven by a potential of the
type $V(\phi)=\lambda^4\exp\left(-\sqrt{\frac{2}{p}}\frac{\phi}{m_{\rm
P}}\right)$. Here $p$ is some number, characterizing the
expansion rate, and $\lambda$ describes the energy scale of
inflation. The corresponding solution of the scale factor in this
scenario is exactly in the power-law form,
$a(t)=\left[\frac{t}{(p+1)l}\right]^p$, which is the reason for this scenario's
name ($\lambda$ is re-parameterized as $l$ in $a(t)$).

In this model, an additional exit mechanism is needed, since the model
itself cannot stop inflation. Usually, we assume that such a
mechanism exists, and the cosmological perturbations are not affected
by this mechanism. The ordinary power-law inflation model in
commutative space-time has been ruled out at about 3$\sigma$ level
by the current Planck data~\cite{Planck22}. Now we explore the noncommutative case.

The power spectrum, the scalar spectral index, the running of the spectral index,
and the tensor-to-scalar ratio were calculated in detail in
Ref.~\cite{HuangLi},
\begin{eqnarray}
\mathcal{P}_\mathcal{R}&=&Ak^{-\frac{2}{p-1}}\left[1-\sigma\left(\frac{k_{\rm c}}{k}\right)^{\frac{4}{p-1}}\right], \label{Ppower}\\
s=n_{\rm s}-1&=&-\frac{2}{p-1}+\frac{4\sigma}{p-1}\left(\frac{k_{\rm c}}{k}\right)^{\frac{4}{p-1}}, \label{spower}\\
\alpha_{\rm s}&=&-\frac{16\sigma}{(p-1)^2}\left(\frac{k_{\rm c}}{k}\right)^{\frac{4}{p-1}}, \label{alphapower}\\
r&=&\frac{16}{p}, \label{rpower}
\end{eqnarray}
where
\begin{eqnarray}
A&=&\left[\frac{(2p-1)p}{(p+1)^2l^2}\right]^{\frac{p}{p-1}}\frac{pl_{\rm P}^2}{8\pi^2}, \label{A}\\
k_{\rm c}&=&\left[\frac{(2p-1)p}{(p+1)^2}\right]^{\frac{p+1}{4}}\frac{1}{l_{\rm s}}\left(\frac{l_{\rm s}}{l}\right)^p, \label{kc}\\
\sigma&=&\frac{4(p-2)(2p+1)p^2}{(p-1)(2p-1)(p+1)^2}, \label{para}
\end{eqnarray}
with $l_{\rm P}=m^{-1}_{\rm P}=8.10\times 10^{-35}~{\rm
m}=2.62\times 10^{-57}$~Mpc being the Planck scale. From
Eqs.~(\ref{Ppower})--(\ref{alphapower}), we find that
$\mathcal{P}_\mathcal{R}$, $s$, and $\alpha_{\rm s}$ are the
functions of three parameters, $A$, $k_{\rm c}$, and $\sigma$, and
further of another three parameters, $p$, $l$, and $l_{\rm s}$.
Thus, with the joint Planck+WP+BAO results~\cite{Planck22} for
$\mathcal{P}_\mathcal{R}$, $s$, and $\alpha_{\rm s}$, we are able to
constrain the allowed ranges of all these parameters.

From Eqs.~(\ref{spower}) and (\ref{alphapower}), we get
$s=-\frac{2}{p-1}-\frac{p-1}{4}\alpha_{\rm s}$, so
\begin{eqnarray}
p=1-\frac{2}{\alpha_{\rm s}}\left(\sqrt{s^2-2\alpha_{\rm s}}+s\right). \n
\end{eqnarray}
For their central values in Eq.~(\ref{salpha}), $s=-0.0393$ and $\alpha_{\rm s}=-0.021$, we obtain
\begin{eqnarray}
p=17.1. \n
\end{eqnarray}
As a consequence, from Eq.~(\ref{rpower}), we have
\begin{eqnarray}
r=0.936. \n
\end{eqnarray}
Hence, we again find an overlarge tensor-to-scalar ratio in the
power-law inflation model in noncommutative space-time. In other words,
we say that the scale factor grows not as fast as
needed in this model, since its power should be $p>16/0.25=64\gg 17.1$, if
$r<0.25$~\cite{Planck22}.

Below we give all the other relevant parameters, $\sigma$, $k_{\rm
c}$, $A$, $l$, $l_{\rm s}$, and $M_{\rm s}$. For $p=17.1$, from
Eq.~(\ref{para}), we find
\begin{eqnarray}
\sigma=3.55. \n
\end{eqnarray}
Thus, from Eq.~(\ref{alphapower}), we have
\begin{eqnarray}
k_{\rm c}=3.98\times 10^{-6}~{\rm Mpc}^{-1}, \n
\end{eqnarray}
where we pick the central value $\alpha_{\rm s}=-0.021$ and the
pivot scale $k_{\ast}=0.05~{\rm Mpc}^{-1}$~\cite{Planck22}.
Moreover, from Eq.~(\ref{Ppower}), we obtain
\begin{eqnarray}
A=2.29\times 10^{-9}~{\rm Mpc}^{-0.124}, \n
\end{eqnarray}
where $\mathcal{P}_\mathcal{R}(k_\ast=0.05~{\rm
Mpc}^{-1})=2.196\times 10^{-9}$~\cite{Planck16}. Furthermore, from
Eq.~(\ref{A}), we get
\begin{eqnarray}
l=1.23\times 10^{-27}~{\rm m}. \n
\end{eqnarray}
Finally, from Eq.~(\ref{kc}), we arrive at
\begin{eqnarray}
l_{\rm s}&=&4.15\times 10^{-31}~{\rm m}=5.13\times 10^3l_{\rm P}, \n\\
M_{\rm s}&=&4.76\times 10^{14}~{\rm GeV}=1.95\times 10^{-4}m_{\rm P}. \n
\end{eqnarray}
We should state that all the values above are taken as the central
values, extracted from the joint Planck+WP+BAO
data~\cite{Planck16,Planck22}. It is unnecessary to
discuss the specific meaning of each parameter, since it is only
the trouble of the large tensor-to-scalar ratio that is already
enough for us to discard the noncommutative power-law inflation
model.

\section{Discussion and Summary} \label{sec:sum}

In this paper, we reexamine the inflationary models in
noncommutative space-time, with the constraints from the latest
observational data from the Planck mission~\cite{Planck16,Planck22}.
Firstly, we demonstrate in general that the noncommutative effects
can help to induce nontrivial modifications to the primordial
parameters. Then, as examples, we explore the possible
noncommutative effects in two specific noncommutative inflation
scenarios---the chaotic and power-law inflation models---but find
incurable discrepancies between the predictions from these models
and the joint Planck+WP+BAO data. First, if the observational
constraints of the spectral index and the running of the spectral
index, $n_{\rm s}=0.9607\pm 0.0063$ and $\alpha_{\rm
s}=-0.021^{+0.012}_{-0.010}$ ($68\%$ CL)~\cite{Planck22}, are
satisfied, the noncommutative effects will always produce a too-large tensor-to-scalar ratio: $r>0.41$ for the $\phi^2$ inflation
model, $r>0.51$ for the $\phi^4$ inflation model, and $r=0.936$ (in
this case we only take the best-fit values as an example) for the
power-law inflation. Furthermore, if we let the observational upper
bound for the tensor-to-scalar ratio, $r<0.25$~\cite{Planck22}, be
satisfied, we find that the running of the spectral index
$\alpha_{\rm s}$ will be tiny, still around ${\cal O}(10^{-3})$,
i.e., under such circumstances the noncommutative effects cannot
provide considerable help in generating large running spectral
index. For instance, the maximum of $\alpha_{\rm s}$ is $-0.0025$ in
the noncommutative $\phi^2$ inflation model, which is much smaller
than its central value $-0.021$ from the joint Planck+WP+BAO
data~\cite{Planck22}. Last, we find that, even though we give up
the aim of producing a large running of the spectral index, the noncommutative
effects do not provide a better fit to the data but make the
situation worse. For example, considering the $r$--$n_{\rm s}$
relationship, we find that the noncommutative effects give rise to a
larger tensor-to-scalar ratio, if the spectral index is fixed; as a
result, the $\phi^2$ inflation model with the noncommutative
parameter $\mu\gtrsim 0.05$ is ruled out at the $95\%$ CL by the
joint Planck+WP+BAO data. Actually, the same conclusion would be drawn
if we use the SPT and ACT results to make a similar analysis.
Altogether, we can conclude for safety
that the inflationary models in noncommutative space-time are
disfavored by the current CMB data.

\begin{acknowledgments}
We are very grateful to Yun-He Li and Dominik J. Schwarz for helpful discussions. NL is
supported by the National Natural Science Foundation of China (Grant
No.~11105026). XZ is supported by the National Natural Science
Foundation of China (Grants No.~10705041, No.~10975032, and No.~11175042)
and by the National Ministry of Education of China (Grants
No.~NCET-09-0276 and No.~N120505003).
\end{acknowledgments}


\begin{thebibliography}{99}
\bibitem{Guth}
A. Guth, Phys. Rev. D {\bf 23}, 347 (1981).

\bibitem{Linde}
A. Linde, Phys. Lett. B {\bf 108}, 389 (1982).

\bibitem{Steinhardt}
A. Albrecht and P. Steinhardt, Phys. Rev. Lett. {\bf 48}, 1220 (1982).

\bibitem{WMAP}
G. Hinshaw {\it et al.}, arXiv:1212.5226 [astro-ph.CO].

\bibitem{Harrison}
E. R. Harrison, Phys. Rev. D {\bf 1}, 2726 (1970).

\bibitem{Zel'dovich}
Ya. B. Zel'dovich, Mon. Not. Roy. Astron. Soc. {\bf 160}, 1 (1972).

\bibitem{Planck16}
P. A. R. Ade {\it et al.}, 
arXiv:1303.5076 [astro-ph.CO]. 

\bibitem{Planck22}
P. A. R. Ade {\it et al.}, 
arXiv:1303.5082 [astro-ph.CO]. 

\bibitem{Brandenberger}
R. Brandenberger and P. M. Ho, Phys. Rev. D {\bf 66}, 023517 (2002) [arXiv:hep-th/0203119]. 

\bibitem{HuangLi1}
Q. G. Huang and M. Li, J. High Energy Phys. {\bf 06}, 014 (2003) [arXiv:hep-th/0304203]. 

\bibitem{HuangLi2}
Q. G. Huang and M. Li, J. Cosmol. Astropart. Phys. {\bf 11}, 001 (2003) [arXiv:astro-ph/0308458]. 

\bibitem{HuangLi}
Q. G. Huang and M. Li, Nucl. Phys. B {\bf 713}, 219 (2005) [arXiv:astro-ph/0311378]. 

\bibitem{Huang}
Q. G. Huang and M. Li, Nucl. Phys. B {\bf 755}, 286 (2006) [arXiv:astro-ph/0603782]. 

\bibitem{Zhang:2006ta}
X. Zhang and F. Q. Wu, Phys. Lett. B {\bf 638}, 396 (2006)
[arXiv:astro-ph/0604195].

\bibitem{Huang:2006zu}
Q. G. Huang, Phys Rev. D {\bf 74}, 063513 (2006)
[arXiv:astro-ph/0605442].

\bibitem{Zhang:2006cc}
X. Zhang, J. Cosmol. Astropart. Phys. {\bf 12}, 002 (2006) [arXiv:hep-th/0608207].

\bibitem{Cai:2007xr}
  Y. F.~Cai and Y. S.~Piao,
  Phys.\ Lett.\ B {\bf 657}, 1 (2007)
  [arXiv:gr-qc/0701114].

\bibitem{Cai:2007et}
  Y. F.~Cai and Y.~Wang,
  J. Cosmol. Astropart. Phys. {\bf 06}, 022 (2007)
  [arXiv:0706.0572 [hep-th]].


\bibitem{Xue:2007bb}
  W.~Xue, B.~Chen, and Y.~Wang,
  J. Cosmol. Astropart. Phys. {\bf 09}, 011 (2007)
  [arXiv:0706.1843 [hep-th]].



\bibitem{wmap3}
D. N. Spergel {\it et al.}, 
Astrophys. J. Suppl. {\bf 170}, 377 (2007) [arXiv:astro-ph/0603449].

\bibitem{Yoneya}
T. Yoneya, in {\it Wandering in the Fields}, edited by K. Kawarabayashi and A. Ukawa (World Scientific, Singapore, 1987), p. 419.

\bibitem{Li}
M. Li and T. Yoneya, Phys. Rev. Lett. {\bf 78}, 1219 (1997) [arXiv:hep-th/9611072]. 

\bibitem{KKLMMT}
S. Kachru {\it et al.}, J. Cosmol. Astropart. Phys. {\bf 10} 013 (2003) [arXiv:hep-th/0308055].

\bibitem{Huang:2006ra}
Q. G. Huang, M. Li, and J. H. She, J. Cosmol. Astropart. Phys. {\bf
11}, 010 (2006) [arXiv:hep-th/0604186].

\bibitem{Ma:2008rf}
Y. Z. Ma and X. Zhang, J. Cosmol. Astropart. Phys. {\bf 03}, 006
(2009) [arXiv:0812.3421 [astro-ph.CO]].

\bibitem{Ma:2013xma}
Y. Z. Ma, Q. G. Huang, and X. Zhang, Phys. Rev. D {\bf 87}, 103516 (2013) [arXiv:1303.6244 [astro-ph.CO]].

\bibitem{SPT}
  Z.~Hou, C.~L.~Reichardt, and K.~T.~Story {\it et al.},
  arXiv:1212.6267 [astro-ph.CO].

\bibitem{ACT}
  J.~L.~Sievers, R.~A.~Hlozek, and M.~R.~Nolta {\it et al.},
  arXiv:1301.0824 [astro-ph.CO].

\bibitem{Lindechaotic}
A. Linde, Phys. Lett. B {\bf 129}, 177 (1983). 

\bibitem{powerlaw}
F. Lucchin and S. Matarrese, Phys. Rev. D {\bf 32}, 1316 (1985). 

\end{thebibliography}
\end{document}